\documentclass[aip,amsmath,amssymb,reprint]{revtex4-1}
\usepackage{graphicx}
\usepackage{dcolumn}
\usepackage{bm}
\usepackage[utf8]{inputenc}
\usepackage[T1]{fontenc}
\usepackage{mathptmx}
\renewcommand{\vec}[1]{{\mathrm{\mathbf{ #1}}}}

\begin{document}

\title[Learning on-top: regressing the on-top pair density for real-space visualization of electron correlation]{Learning on-top: regressing the on-top pair density for real-space visualization of electron correlation}

\author{Alberto Fabrizio}
\thanks{ The authors contributed equally to this work. }
\affiliation{ Laboratory for Computational Molecular Design, Institute of Chemical Sciences and Engineering, \'{E}cole Polytechnique F\'{e}d\'{e}rale de Lausanne, CH-1015 Lausanne, Switzerland
}%
\affiliation{ National Centre for Computational Design and Discovery of Novel Materials (MARVEL), \'{E}cole Polytechnique F\'{e}d\'{e}rale de Lausanne, CH-1015 Lausanne, Switzerland
}%
\author{Ksenia R. Briling}%
\thanks{ The authors contributed equally to this work. }
\affiliation{ Laboratory for Computational Molecular Design, Institute of Chemical Sciences and Engineering, \'{E}cole Polytechnique F\'{e}d\'{e}rale de Lausanne, CH-1015 Lausanne, Switzerland
}%
\author{David D. Girardier}%
 \affiliation{ Laboratory for Computational Molecular Design, Institute of Chemical Sciences and Engineering, \'{E}cole Polytechnique F\'{e}d\'{e}rale de Lausanne, CH-1015 Lausanne, Switzerland
}%
\author{Clemence Corminboeuf}
\email{clemence.corminboeuf@epfl.ch}
\affiliation{ Laboratory for Computational Molecular Design, Institute of Chemical Sciences and Engineering, \'{E}cole Polytechnique F\'{e}d\'{e}rale de Lausanne, CH-1015 Lausanne, Switzerland
}%
\affiliation{ National Centre for Computational Design and Discovery of Novel Materials (MARVEL), \'{E}cole Polytechnique F\'{e}d\'{e}rale de Lausanne, CH-1015 Lausanne, Switzerland
}%

\date{\today}

\begin{abstract}

The on-top pair density [$\Pi(\vec{r})$] is a local quantum-chemical property, which reflects the probability of two electrons of any spin to occupy the same position in space. Being the simplest quantity related to the two-particles density matrix, the on-top pair density is a powerful indicator of electron correlation effects and, as such, it has been extensively used to combine density functional theory and multireference wavefunction theory. The widespread application of $\Pi(\vec{r})$ is currently hindered by the need for post-Hartree--Fock or multireference computations for its accurate evaluation. In this work, we propose the construction of a machine learning model capable of predicting the CASSCF-quality on-top pair density of a molecule only from its structure and composition. Our model, trained on the GDB11-AD-3165 database, is able to predict with minimal error the on-top pair density of organic molecules bypassing completely the need for \textit{ab initio} computations. The accuracy of the regression is demonstrated using the on-top ratio as a visual metric of electron correlation effects and bond-breaking in real-space. In addition, we report the construction of a specialized basis set, built to fit the on-top pair density in a single, atom-centered expansion. This basis, cornerstone of the regression, could be potentially used also in the same spirit of the resolution-of-the-identity approximation for the electron density.

\end{abstract}

\maketitle

\section{\label{sec:level1}Introduction}

The spinless on-top pair density [OTPD, $\Pi(\vec{r})$, Eq. \ref{eq:pi}] is a local quantum-chemical quantity, which represents the probability density for two electrons of any spin to occupy the same position in real-space:\cite{mcweeny1992methods}
\begin{equation}
\label{eq:pi}
        \Pi(\vec{r}) = \binom{N}{2} \int \left| \Psi(\vec{x}_1,...\vec{x}_N) \right|^2 d \sigma_1 ... d\sigma_N d\vec{r}_3 ... d\vec{r}_N |_{\vec{r}_1=\vec{r}_2=\vec{r}},
\end{equation}
where $\sigma$ is the spin variable, $\vec{r}$ is the space variable and $\vec{x}=(\vec{r},\sigma)$.

Because the on-top pair density is the simplest local property that is still related to the two-particle density matrix (2-PDM), $\Pi(\vec{r})$ occupies a prominent place in the history of electron correlation in density functional functional theory. Already in the mid-90s it was shown that the on-top pair density is the fundamental quantity that justifies the accurate results obtained on some multireference systems with broken-symmetry LSDA and GGA computations.\cite{Perdew1995} Indeed, chemical situations such as the stretched $\mathrm{H_2}$ molecule suffer from an apparent dilemma (akin to L\"{o}wdin's symmetry dilemma in Hartree--Fock):\cite{Lowdin1955} preserving the correct spin distribution yields inaccurate electronic energies, and viceversa.\cite{Gunnarsson1976, Gorling1993} Concrete examples of this problem have been already known since the late 70s when it was demonstrated that the local spin-density approximation (LSDA) is only justified for single-determinant states and that the local spin density philosophy is not compatible with the spin densities of multireference wavefunctions.\cite{Ziegler1977} In contrast, Perdew, Ernzerhof, Savin and Burke showed that despite yielding incorrect spin distribution, broken-symmetry solutions already at the LSDA level mimicked very closely the on-top pair density computed with multireference wavefunction theory.\cite{Perdew1995,Perdew1997} It was later argued that the ability of LSDA to produce satisfyingly accurate spin-independent quantities [$\rho(\vec{r})$ and $\Pi(\vec{r})$] is the reason beyond the unexpectedly fair behavior in many chemical situations of this otherwise very crude approximation to exact DFT.\cite{Burke1998}

Besides its usefulness to re-interpret the results of approximate functionals, the on-top pair density has been traditionally used to combine wavefunction theory and DFT. As early as the 1991, the pioneering work of Moscard\'{o} and San-Fabi\'{a}n\cite{Moscardo1991,Moscardo1991a} showed that it is possible to use the on-top pair density from a configuration-interaction wavefunction (CID) to help density functionals to capture the intrinsic two-body nature of the correlation energy. This first result only slightly anticipated the work of Becke, Savin and Stoll who proposed a complete re-evaluation of the local spin-density approach and replaced the spin densities [$\rho_\alpha (\vec{r})$, $\rho_\beta (\vec{r})$] as independent variables with $\rho(\vec{r})$ and $\Pi(\vec{r})$.\cite{Becke1995} Since the electron density and the on-top pair density can be evaluated straightforwardly from a multiconfigurational wavefunction, the proposed substitution allows extending the machinery of Kohn--Sham DFT to multi-determinant reference states.

On the momentum of these works, the on-top pair density has been then mostly applied as ingredient of mixed electronic structure methods such as multiconfiguration DFT (MC-DFT),\cite{Miehlich1997} complete active space DFT (CAS-DFT)\cite{Grafenstein2000,Takeda2004,Gusarov2004,Hapka2020} and, more recently, multiconfiguration pair-density functional theory (MC-PDFT) by Truhlar, Gagliardi and coworkers.\cite{LiManni2014,Gagliardi2017,Sand2018} In these latter works, at least part of the total electronic energy is obtained non-variationally through a density functional, whose input is or directly derives from the on-top pair density of a converged multiconfigurational wavefunction. Among these methods, MC-PDFT has emerged as one of the closest realizations of the original suggestion to combine multireference wavefunction theory and DFT through the on-top pair density.\cite{Becke1995} In the last few years, this framework has been shown to produce benchmark-quality results for typically challenging electronic structures,\cite{Bao2018} including charge-transfer complexes,\cite{Ghosh2015} main-group and transition metal thermochemistry,\cite{Carlson2015,Bao2017} barrier heights,\cite{Carlson2015} and electronic excitations.\cite{Hoyer2016} Complementary to the applications of the on-top pair density for method development, the success of MC-PDFT fueled a renewed interest in its potential application as a visualization tool, especially in the form of the on-top ratio [\textit{i.e.} the ratio between the $\Pi(\vec{r})$ and $\rho^2(\vec{r})/4$] and its partial derivatives.\cite{Carlson2019} These quantities are effective real-space metrics to identify and visualize the effects of electron correlation, as well as to characterize fundamentally the different types of bonds and covalent-bond breaking.\cite{Carlson2017,Carlson2019}

While the applications of the on-top pair density demonstrate its key role in the quantitative and qualitative description of electron correlation, its wide-spread use is severely limited by the necessity to perform post-Hartree--Fock or, more commonly, multireference computations to obtain accurate two-particle density matrices. In the last few years, the recourse to artificial intelligence, more precisely to machine learning (ML), has been increasingly proposed as an effective strategy to access fundamental quantum-chemical objects bypassing the bottleneck of computationally demanding methods.\cite{Ramakrishnan2015,Brockherde2017,Bartok2017,Chmiela2018,Welborn2018,Smith2019,Fabregat2020,Qiao2020,husch2020improved} In this context, two of us have recently proposed,\cite{Grisafi2019} improved\cite{Fabrizio-CHEMSCI-2019} and demonstrated the large applicability\cite{Fabrizio-CHEMSCI-2019, Fabrizio2020} of a local machine learning model of the molecular electron density capable of accurately reproducing the complex rotational symmetries of the density in real-space. While applied to date only for the regression of the ground-state electron density, the architecture of the framework is general and allows, in principle, to target any local field as long as its decomposition onto an atom-centered basis set is possible.

In this work, we adapt the electron density learning framework to build a direct mapping between the structure and composition of a molecule and its CASSCF-quality on-top pair density. Granting ready access to the information contained in $\Pi(\vec{r})$, we conceive our model as an effective tool for the real-space visualization of electron correlation effects and covalent-bond breaking. To achieve this goal, the framework is built on the recently introduced GDB11-AD-3165 database of Kulik and coworkers,\cite{Duan2020} which consists of 3165 small organic molecules both at their equilibrium and distorted geometries. The GDB11-AD-3165 set has not only the advantage of hosting inherently multiconfigurational molecules and structures, but it also reports accurate CASSCF results, including the weight of the dominant electronic configuration ($|C_0|^2$) and the precise extent of the active spaces used.

\section{\label{sec:level2} The $\phi^{\mathrm{OTPD}}(\vec{r})$ specialized basis set}

 \begin{figure*}[!htb]
    \centering
    \includegraphics[width=0.75\textwidth]{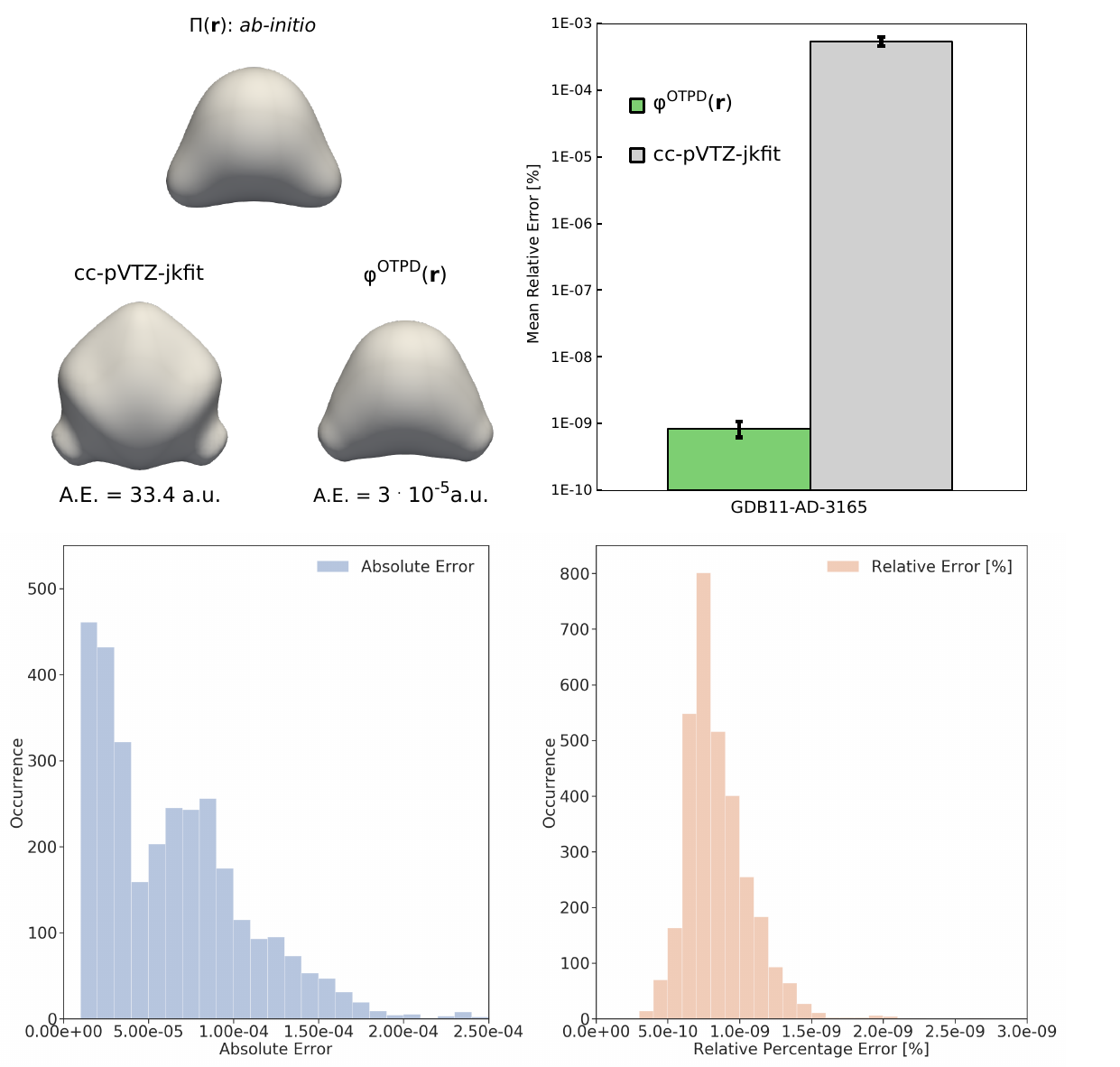}
    \caption{\emph{(top, left)} Qualitative comparison of fitting performance on the water molecule; isosurfaces: 0.005 $\mathrm{e^2 \cdot Bohr^{-6}}$. \emph{(top, right)} Comparison between the $\phi^{\mathrm{OTPD}}(\vec{r})$ and the cc-pVTZ-jkfit\cite{Weigend2002} (RI-basis) for the fitting of the on-top pair density (CASSCF/cc-pvTZ) on the GDB11-AD-3165 set.\cite{Duan2020} Height of the histogram correspond to the mean relative percentage error on the GDB11-AD-3165.  \emph{(bottom)} Distribution of the absolute and relative [\%] fitting error for $\phi^{\mathrm{OTPD}}(\vec{r})$ on the GDB11-AD-3165 set.}
    \label{fig:decoFULLSet}
\end{figure*}

The symmetry-adapted Gaussian process regression framework (SA-GPR) used in this work was originally proposed by Cs{\'{a}}nyi, Ceriotti and coworkers\cite{Grisafi2018} and is a generalization of traditional GPR, capable of encoding the complex symmetries of Cartesian and spherical tensors of any order through a hierarchy of kernels. As already demonstrated in the case of the ground-state electron density,\cite{Grisafi2019, Fabrizio-CHEMSCI-2019} SA-GPR captures accurately the rotational symmetries of a local scalar field, subject to the constraint that its angular dependence is described by a single set of atom-centered spherical harmonics [$Y_l^m(\theta, \phi)$]. In this way, any field can be decomposed into a sum of spherical tensor components with rank $l$, beginning with the spherically symmetric elements ($l=0$) and increasingly adding anisotropy ($l>0$). The treatment of the radial part of the field is subordinated to a much lighter constraint, as the only requirement is for the chosen atom-centered basis functions to have a similar long-range ($r \to \infty$) asymptotic behavior.

The local decomposition strategy is not only an efficient mathematical expedient to represent a scalar field and ensure its compatibility with SA-GPR, but it is also the foundation of the scalability and transferability of the model. These two last properties stem from the combination of the locality of the decomposed field with the locality of the molecular representation used within the SA-GPR framework (symmetry-adapted smooth overlap of atomic positions, $\lambda$-SOAP) and allow the linear-scaling prediction of the target property for a complex molecule, while restraining the training procedure only to much smaller fragments. This type of Lego approach is very similar in spirit to well-established linear scaling techniques in quantum chemistry such as Mezey's molecular electron density Lego
assembler (MEDLA)\cite{Walker1993,Walker1994} and adjustable density matrix assembler (ADMA).\cite{Exner2002,Exner2003,Szekeres2005}

For these reasons, we seek a decomposition of the on-top pair density [$\Pi(\vec{r})$] in the form of
\begin{equation}
\label{eq:deco}
    \Pi(\vec{r}) = \frac{1}{2} \sum_{abcd}^{N_{\rm AO}} D_{abcd}\, \chi_a(\vec{r}) \chi_b(\vec{r}) \chi_c(\vec{r}) \chi_d(\vec{r}) = \sum_i c_i \phi_i(\vec{r}),
\end{equation}
where $D_{abcd}$ is the two-particle density matrix at any chosen level of theory (\textit{e.g.} HF, DFT, CASSCF, CCSD(T), \textit{etc.}), $\chi(\vec{r})$ are atomic orbitals, $c_i$ are decomposition expansion coefficients [$i=\{\mathrm{atom},n,l,m\}$] and $\phi_i(\vec{r})$ are a set of suitable basis functions for the decomposition.

The expansion onto the non-orthogonal basis shown in rightmost side of Eq. \ref{eq:deco} has to be interpreted in the same spirit as the well-established density-fitting approximation (RI or resolution-of-the-identity) in the context of the electron density. We have already demonstrated that the use of specialized basis sets (auxiliary or RI-basis) optimized for the density-fitting approximation improves dramatically the regression of the electron density within the symmetry-adapted framework. However, as reported for the show-case water molecule in the top, left panel of Figure \ref{fig:decoFULLSet}, common RI-basis functions result in unacceptable errors in the decomposition the on-top pair density, introducing instabilities in the shape of spurious anisotropic features in the field.

To better understand the reason behind the spectacular failure of RI-basis sets for the fitting of the on-top pair density, it is sufficient to consider the form of $\Pi(\vec{r})$ at the Hartree--Fock level:\cite{Becke1995}
\begin{equation}
\label{eq:PiHF}
\begin{split}
    \Pi^{\rm HF}(\vec{r}) & = \frac{1}{2} \sum_{abcd}^{N_{\rm AO}} [D_{ab} D_{cd} - \tfrac{1}{2}D_{ad} D_{cb}] \chi_a(\vec{r}) \chi_b(\vec{r}) \chi_c(\vec{r}) \chi_d(\vec{r}) \\
    & = \frac{\rho^2(\vec{r})}{4},
\end{split}
\end{equation}
where $D_{ab}$ is the Hartree--Fock one-particle density matrix and $\rho(\vec{r})$ the electron density. Equation \ref{eq:PiHF} shows that, even excluding any effect due to electron correlation, the on-top pair density has higher amplitudes near the nuclei and decays much faster (twice as fast) than electron density, for which the RI-bases were optimized. Unfortunately, simple solutions such as doubling the exponents of the RI-basis improves only marginally the fitting error (absolute error = 5.8 a.u. on the water molecule; standard RI-basis absolute error = 33.5 a.u.).

For this reason, we propose here the construction of a specialized basis [$\phi^{\rm OTPD}(\vec{r})$], adapted to reduce the fitting error on the on-top pair density. Inspired by the work of Neese and coworkers,\cite{Stoychev2017} we divide the optimization procedure in two distinct phases. A first set of basis exponents ($\alpha$) for each angular momentum (up to $l=4$) and for each element in the GDB11-AD-3165 set (H, C, N, O) is generated by optimization of the three coefficients ($\alpha_0$, $\beta$, and $\gamma$) of the well-tempered formula:\cite{Huzinaga1990}

\begin{equation}
    \label{eq:wellT}
    \alpha_i = \beta \alpha_{i-1} \left( 1+ \gamma \left( \frac{i-2}{N+1} \right)^2 \right),\quad i = 2, \ldots, N
\end{equation}

The optimization is performed to minimize the fitting error of the on-top pair density at HF/cc-pVTZ of $\mathrm{H_2}$, $\mathrm{H_2O}$, $\mathrm{H_2O_2}$, $\mathrm{C_2H_2}$, $\mathrm{C_2H_4}$, $\mathrm{C_2H_6}$, $\mathrm{CH_4}$, $\mathrm{NH_3}$, $\mathrm{N_2H_2}$, and $\mathrm{N_2}$ (the geometries were taken from the original set by Ahlrichs and coworkers).\cite{Weigend2005} This first crude optimization was then used as an initial condition for the complete relaxation of each exponent with analytical gradients using the BFGS\cite{B1970,F1970,G1970,S1970} algorithm. Further details about $\phi^{\mathrm{OTPD}}(\vec{r})$ and its construction are reported in the Supplementary Material.

As the on-top pair density is a powerful indicator of static (or left-right,\cite{Handy_Mol.Phys.2001} or non-dynamical)\cite{Mok1996} electron correlation, we choose to build our model on the recently published GDB11-AD-3165 dataset.\cite{Duan2020} Figure \ref{fig:decoFULLSet} shows the comparison of the decomposition error on GDB11-AD-3165 between the optimized $\phi^{\mathrm{OTPD}}(\vec{r})$ and the standard cc-pVTZ-jkfit\cite{Weigend2002} basis. The on-top pair densities were obtained at the CASSCF/cc-pVTZ level using the same active spaces as prescribed in the original reference.\cite{Duan2020}

In the figure, the fitting error of $\phi^{\mathrm{OTPD}}(\vec{r})$ is reported using two distinct integral metrics: the absolute error
\[
\mathrm{A.E.} = \int d\vec{r} \big| \Pi(\vec{r}) - \sum_i c_i \phi^{\mathrm{fitting}}(\vec{r}) \big|^2
\]
and the relative error
\[
\mathrm{R.E.} = \int d\vec{r} \big| \Pi(\vec{r}) - \sum_i c_i \phi^{\mathrm{fitting}}(\vec{r}) \big|^2 \bigg/ \int d\vec{r}\, \Pi^2(\vec{r}).
\]
The absolute error is extremely sensitive to the inaccuracies of the fitting near the nuclei and, generally, in the regions with high OTPD amplitudes. On the other hand, the relative error is a more fair metric since, in contrast to the absolute error, it does not depend on the system size. Regardless of the metric chosen, however, $\phi^{\mathrm{OTPD}}(\vec{r})$ is orders of magnitudes more accurate for the fitting of the on-top pair density than standard density-fitting basis sets of comparable size. From a chemical perspective, the GDB11-AD-3165 is divisible between compounds with marked multiconfigurational character (often distorted structures) and single-reference molecule (often at the equilibrium geometry). While this distinction exist in the dataset, the fitting error does not depend on the character of the molecular ground-state (see Supplementary Material for further details).  Since the fitting error persists also after the learning procedure, reaching a high accuracy in the decomposition step is the cornerstone of any application of the machine learned on-top pair density.

\section{\label{sec:level3}Learning curves}

The optimization of the specialized $\phi^{\mathrm{OTPD}}(\vec{r})$ allows the efficient projection of the on-top pair density onto a non-orthogonal, atom-centered (local) basis and promotes its compatibility with the symmetry-adapted Gaussian process regression (SA-GPR) framework. To train the model and assess its accuracy, the GDB11-AD-3165 dataset was randomly split into a training set of 2550 molecules and a test set containing the remaining 615 compounds ($\sim20$~\% of the total). Because of its size and its chemical diversity, the GDB11-AD-3165 database contains a tremendous number of atom-centered environments ($\chi\sim36\,000$). On the other hand, most of these local environments are very similar to each other and are redundant for the regression purposes. The locality of the molecular representation combined with the relative redundancy of chemistry at a short, atom-centered cutoff allows reducing significantly the computational cost of the regression by only choosing a subset of the $M$ most different environments. Akin to our previous work on the electron density, we set $M=1000$ and perform a farthest point sampling in the environment space to form a basis for the regression of the on-top pair density.

\begin{figure}[!htb]
    \centering
    \includegraphics[width=0.45\textwidth]{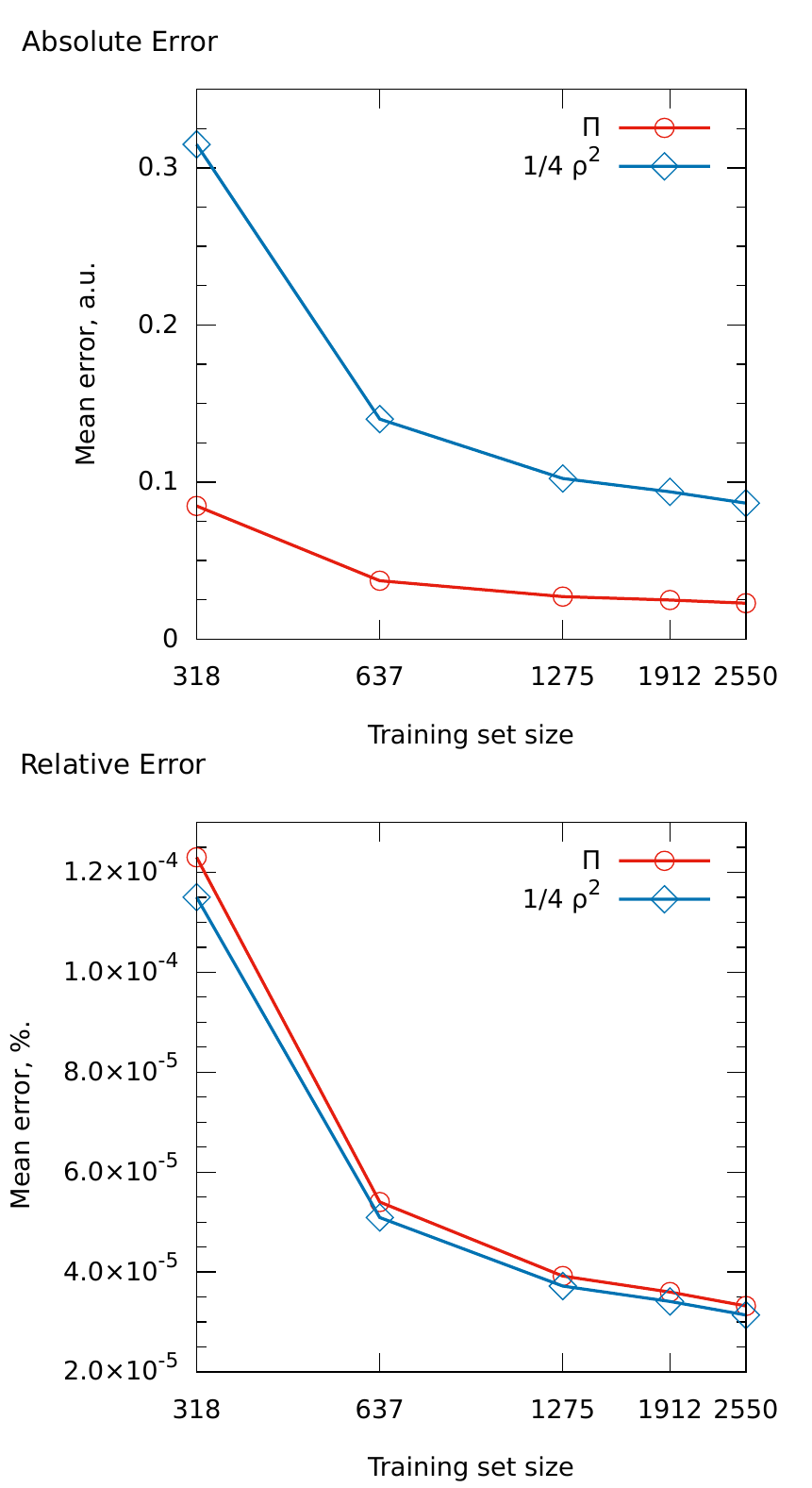}
    \caption{Learning curves with respect to fitted on-top pair densities and squared half-densities (ML error). The learning curves of $(\rho(\vec{r})/2)^2$ are shown for comparison purposes. \emph{(top)} Mean absolute error (MAE), \emph{(bottom)} mean relative percentage error on the test set (615 molecules). The color code distinguishes the on-top pair density (red) from the squared half-densities (blue).}
    \label{fig:LC}
\end{figure}

The performance of the regression model, in terms of mean absolute and relative errors (as defined in Section \ref{sec:level2}) is summarized in Figure \ref{fig:LC}. Besides the on-top pair density, the learning curve of the squared half-densities [$(\rho(\vec{r})/2)^2$] are reported for comparison.

In the limit where the on-top pair density is computed at Hartree--Fock level, $\Pi(\vec{r})$ and $(\rho(\vec{r})/2)^2$ has the same value at every point of space and their learning curve should overlap perfectly. As reported in upper panel of Figure \ref{fig:LC}, however, the absolute error of the two fields differ significantly throughout the learning. In reality the difficulty of the learning exercise is very similar for the two fields as shown using the relative percentage error. The difference between the two metrics is the direct manifestation of an underlying physical effect: the presence of electron correlation at CASSCF level decreases in average the amplitudes of $\Pi(\vec{r})$ with respect to those of independent particles.

Overall, the performance of the model on the test set is more than satisfying with a mean relative error at the full training set (2550 molecules) of only $\mathrm{3.32\cdot 10^{-5}~\%}$ for $\Pi(\vec{r})$ and $\mathrm{3.14\cdot 10^{-5}~\%}$ for $(\rho(\vec{r})/2)^2$. These errors refer to the deviation from the fitted fields and, thus, represent the machine learning error only. Nevertheless, the fitting error with respect to the \textit{ab initio} reference (see Section \ref{sec:level2}) is orders of magnitudes lower than the prediction error, which allows the direct comparison between prediction and \textit{ab initio} results.

\begin{figure}[!htb]
    \centering
    \includegraphics[width=0.45\textwidth]{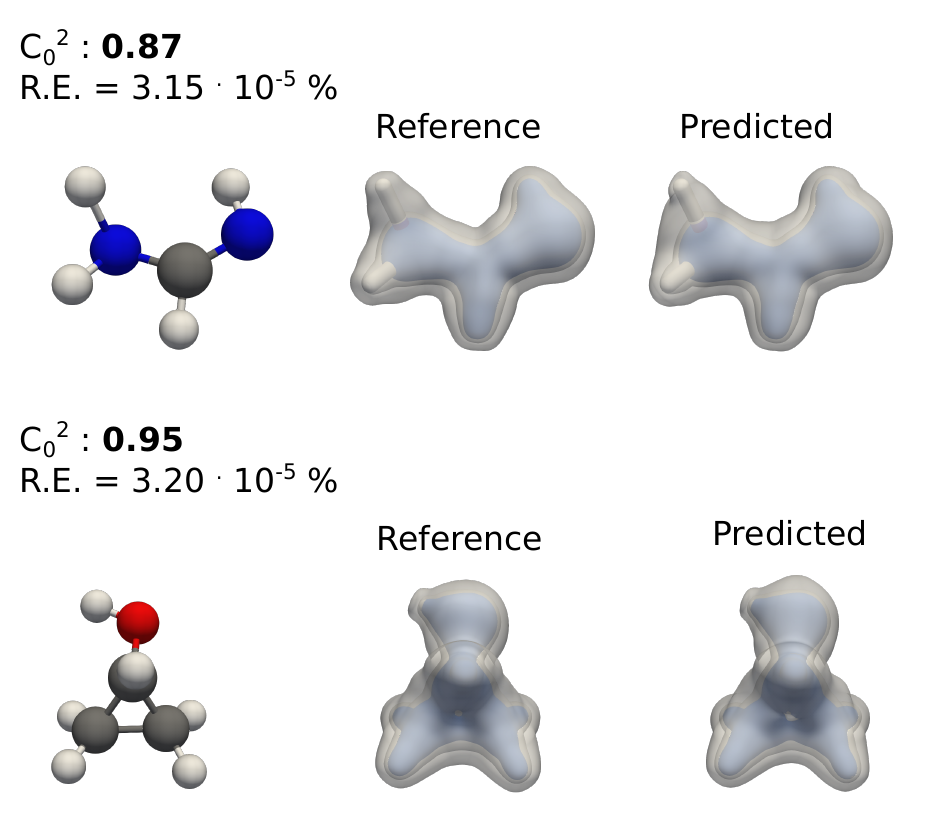}
    \caption{Predicted on-top pair density of two illustrative molecules from test set molecules. The reference field (target of the learning) is $\Pi(\vec{r})$ fitted on $\phi^{\mathrm{OTPD}}(\vec{r})$. Overall, the main real-space features and shape of the on-top pair density is well reproduced by the predictions. Isovalues: 0.01 $\mathrm{e^2 \cdot Bohr^{-6}}$ (blue), 0.005 $\mathrm{e^2 \cdot Bohr^{-6}}$ (innermost, gray), and 0.001 $\mathrm{e^2 \cdot Bohr^{-6}}$ (outermost, gray). The weight of the dominant configuration ($|C_0|^2$) and the relative error of the prediction (R.E.) is reported for each molecule.}
    \label{fig:LC_demo}
\end{figure}

Quantifying the prediction error through a well-defined numerical metric, such as the absolute and relative error, is an objective procedure to estimate the deviation from the targeted fields, but it tells only little about the impact of such an error on the shape of $\Pi(\vec{r})$, its real-space properties and its applicability in methods such as MC-PDFT. For this reason, we report in Figure \ref{fig:LC_demo} two illustrative examples of the quality of predicted on-top pair densities in real-space. The two molecules were selected from the test set following two distinct criteria. First, both molecules are characterized by a relative prediction error similar to the mean on the test set (last point of the learning curve). Second, the first molecule is characterized by an evident multiconfigurational character (CASSCF ${|C_0|^2= 0.87}$), while the electronic structure of the other is already well-described by its most dominant configuration only (${|C_0|^2= 0.95}$).

In both molecules, regardless of their multireference character, the main features and the overall shape of the on-top pair density is well reproduced by the predictions. This is especially true for the highest amplitudes, closer to the nuclei (blue parts in Figure \ref{fig:LC_demo}), while the prediction error is somewhat more evident in the low-$\Pi(\vec{r})$ regions (see \textit{e.g.} the primary amine group of the first molecule). The slight degradation of the quality of the regression further from the nuclei has been already observed in the case of the electron density and it has to be attributed in part to the incompleteness of the basis set and in part to fact that the machine learning algorithm tends to minimize the overall error by reducing the discrepancies in the high-amplitude regions. Overall, the quality of the regression is largely sufficient for our goal to apply the predicted on-top pair densities as real-space metrics, as discussed in the following section.

\section{\label{sec:level4} A visual metric of electron correlation and bond breaking}

The ability to predict from the molecular structure only both the on-top pair densities [$\Pi(\vec{r})$] and squared half-densities [$(\rho(\vec{r})/2)^2$] at CASSCF level promotes their application as real-space metrics for the evaluation of electron correlation effects and the characterization of bond-types and bond-breaking. This kind of application, similar in spirit to other well-established scalar fields based on the electron density,\cite{Bader1991,Bader2007,Johnson2010,Contreras-Garcia2011,DeSilva2014, Grimme2015} has been proposed and meticulously analyzed for simple molecules at CASSCF level by Carlson, Truhlar and Gagliardi.\cite{Carlson2017,Carlson2019} The key quantity to be computed is the on-top ratio, which is defined as:
\begin{equation}
    \label{eq:ontopratio}
    R(\vec{r}) = \frac{\Pi(\vec{r})}{[\rho(\vec{r})/2]^2}
\end{equation}

\begin{figure*}[!htb]
    \centering
    \includegraphics[width=0.75\textwidth]{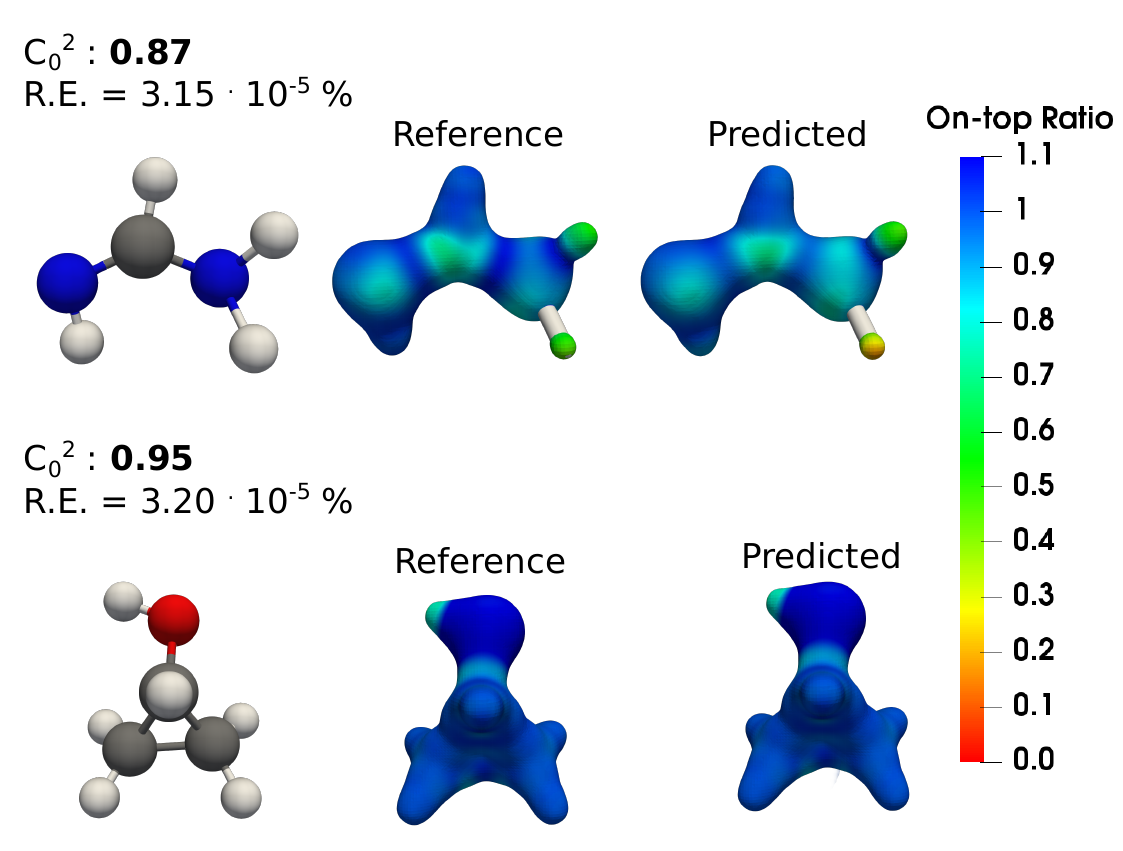}
    \caption{Predicted on-top ratio [color-code, $R(\vec{r})$] for two illustrative molecules of the test set projected on their squared densities (isovalue 0.025 $\mathrm{e^2 \cdot Bohr^{-6}}$). The reference $R$ is computed using the fitted on-top pair densities and squared half-densities.}
    \label{fig:application}
\end{figure*}

Figure \ref{fig:application} reports the on-top ratio computed from the predicted on-top pair densities and squared half-densities of the same two molecules presented in Figure \ref{fig:LC_demo}. For visualization purposes the on-top ratio is projected on the surface of squared half-densities for the two molecules.

Although, from the quantitative perspective, it can be noticed that the on-top ratio from the predicted quantities is slightly underestimated on the hydrogen atom of the first molecule, overall it  reproduces well both the shape and the qualitative distribution of the reference field. More importantly, because of its correct distribution, the chemical interpretation of the predicted on-top ratio is still possible. As shown in the Figure, the on-top ratio is very close to 1 (HF limit) everywhere in the second molecule, where a single electronic configuration dominates the ground-state. The few regions where the ratio decreases slightly (0.85) are a consequence of dynamic electron correlation, which is partially included along with static correlation in the CASSCF computation. In contrast, the first molecule has a significant multiconfigurational character, originating from the significantly elongated N-H bonds in the primary amine (1.24~\AA{} and 1.43~\AA{} respectively, while the standard length is $\sim 1.00$~\AA). In the region of these two hydrogen, the on-top ratio drops significantly and approaches zero. This behavior is not surprising, as at most 1 electron (homolytic bond breaking) would be present on each of the dissociating hydrogens and the $\Pi(\vec{r})$ reflects the probability of two electrons sharing the same position in space. These examples concretely demonstrate how the predicted on-top pair densities and squared half-densities are sufficiently accurate to access valuable real-space information about correlation effects and bond-breaking in molecules.

\section{\label{sec:level5}Conclusion}

The centrality of the on-top pair density for the description of electron correlation in atoms, molecules and materials is well reflected by the immense research effort put in the analysis of its properties and applications. In this work, we have proposed a machine learning model of $\Pi(\vec{r})$ capable of yielding accurate on-top pair densities given the molecular structure only. Bypassing the demanding CASSCF computations and leveraging on its ability to capture the complex symmetries of fields in real-space, our model becomes a powerful tool for the visualization and the analysis of electron correlation effects (both static and dynamic). In principle, the predicted densities and on-top pair densities could be also used as ingredients for electronic structure methods such as MC-PDFT. Nevertheless, this kind of application is substantially limited by the necessity to use the wavefunction to compute the kinetic energy contribution to the total electronic energy.

The regression of $\Pi(\vec{r})$ is also the first concrete demonstration of the generalizability to any real-space field of the SA-GPR framework. While the machine learning machinery remains the same regardless of the chosen target, it is imperative to develop efficient decomposition schemes capable of minimizing the fitting error of the target field onto a single basis set. In this respect, we have reported the construction of a specialized basis set for the fitting of the on-top pair density and any general field which behaves as the square of the electron density.

Taking advantage of its locality, the current model is transferable and allows the regression of the on-top pair density of organic molecules of any size, as long as their local atomic environments are contained within the GDB11-AD-3165 database (for an example, see Supplementary Material). The construction rules of GDB11-AD-3165 (\textit{i.e.} singlet, neutral molecules only with maximum 6 heavy atoms) result into some limitations for the transferability of the current model, since structural patterns such as fused aromatic rings and electronic patterns such as open-shell radicals and charged species are not included. Since the basis set was optimized only for H,C,N,O and the model was trained on a database composed exclusively of organic molecules, transition metal complexes are also not yet compatible with the current form of the framework. Nevertheless, these limitations could be easily lifted in the future by the optimization of specialized basis functions for the transition metal centers and the extension of the training set to any compound of interest, including charged compounds, radicals and transition metal complexes.

As a second advantage of its locality, the model is not restricted to molecules in gas-phase, but it could be applied as a tool to identify and visualize the effects of strong correlation in molecular materials, organic crystals and molecules in solution. This kind of application demands, however, the construction of a training set \textit{ad-hoc}, since it would be crucial for the model to capture the complex effects of intermolecular interactions on $\Pi(\vec{r})$.

\section*{Supplementary Material}

See the Supplementary Material for a detailed description of the construction of $\phi^{\mathrm{OTPD}}(\vec{r})$, the performance of the decomposition scheme on stretched bonds, an application of the framework for a larger molecule (extrapolation), and the numerical data associated to the learning curves reported in the main text.

\begin{acknowledgments}

The authors acknowledge Andrea Grisafi, David M. Wilkins, and Michele Ceriotti for sharing the code to construct the tensorial SOAP kernels. The National Centre of Competence in Research (NCCR) ``Materials' Revolution: Computational Design and Discovery of Novel Materials (MARVEL)'' of the Swiss National Science Foundation (SNSF), the European Research Council (ERC, grant agreement no 817977),
and the EPFL are also acknowledged for financial support.

\end{acknowledgments}

\section*{Data Availability Statement}

The data that support the findings of this study are openly available in Materials Cloud at \url{https://doi.org/10.24435/materialscloud:8z-2p}.

\appendix

\section{Computational Details}

The molecular geometries for all compounds were taken as published in the GDB11-AD-3165 database.\cite{Duan2020} The on-top pair densities and the squared half-densities were computed at CASSCF\cite{Roos1980}/cc-pVTZ\cite{Dunning1989} level as implemented in OpenMolcas.\cite{Fdez.Galvan2019} The size and the composition of the active spaces were taken from the original reference.\cite{Duan2020} The visualization of all the scalar fields and the generation of Figures was performed with a slightly modified version of the Paraview 5.6.0 software.\cite{AHRENS2005} Basis set generation and optimization were performed with an in-house code, which is provided on GitHub (\url{https://github.com/lcmd-epfl/OTPD-basis}).
The projection integrals of the on-top pair density onto the specialized basis set and the basis overlap integrals were computed numerically on a (99, 590) MHL\cite{Murray1993}-Lebedev\cite{Lebedev1992} grid with a crowding factor of 5.0.

The tensorial $\lambda$-SOAP kernels\cite{Grisafi2018} were computed with the following parameters:
environment cutoff $r_\mathrm{cut} = 4$~\AA,
Gaussian smearing $\sigma = 0.3$~\AA,
angular cutoff $l_\mathrm{cut} = 6$,
radial cutoff $n_\mathrm{cut} = 8$,
environmental kernel exponent $\zeta = 2$.
A subset of $M = 1000$ reference environments was taken
to reduce the dimensionality of the regression problem,
and the regularization parameter $\eta$ was set to $10^{-6}$.

\section*{References}
\bibliography{bibliography}

\end{document}


\LARGE {\textbf{Supplementary Material}}

\vspace*{0.75cm}

\large \textbf{Learning on-top: regressing the on-top pair density for real-space visualization of electron correlation}

\vspace*{0.25cm}

Alberto Fabrizio,\textit{$^{a,b}$} Ksenia R. Briling,\textit{$^{a}$} David D. Girardier,\textit{$^{a}$} and Clemence Corminboeuf$^\ast$\textit{$^{a,b}$} \\

\noindent\small{\textit{$^{a}$~Laboratory for Computational Molecular Design, Institute of Chemical Sciences and Engineering, \'Ecole Polytechnique F\'ed\'erale de Lausanne, CH-1015 Lausanne, Switzerland.\\$^{b}$~National Centre for Computational Design and Discovery of Novel Materials (MARVEL), {\'E}cole Polytechnique F{\'e}d{\'e}rale de Lausanne, CH-1015 Lausanne, Switzerland \\ ~\\ E-mail: clemence.corminboeuf@epfl.ch}}

\tableofcontents

\newpage
\section{Construction of \texorpdfstring{$\phi^{\mathrm{OTPD}}(\vec{r})$}{φOTPD(r)}}

The optimization procedure used to construct the specialized basis set proposed in this work consists of two separate stages.\cite{Stoychev2017} At each stage, the fitting error of the on-top pair density is minimized for a given molecule or set of simple compounds, which are representative of the most common oxidation states of the targeted element.\cite{Weigend2005} Because of its ubiquity in organic chemistry, the first element treated was hydrogen, minimizing the fitting error on the closed-shell \ce{H2} molecule at equilibrium geometry. Using the optimized exponents for H, the procedure was repeated for the sets containing carbon (\ce{CH4}, \ce{C2H2}, \ce{C2H4}, and \ce{C2H6}), nitrogen (\ce{N2}, \ce{NH3}, and \ce{N2H2}), and oxygen (\ce{H2O} and \ce{H2O2}).

For each molecule, we search for the set of coefficients $\{ c_i \}$ that approximates the on-top pair density in the least-squares sense,
\begin{equation}
\Pi(\vec r) \approx \sum_i c_i \phi^{\mathrm{OTPD}}_i(\vec r).
\end{equation}
The decomposition coefficients $\vec c$ are
\begin{equation}
\vec c = \vec S^{-1} \vec b,
\end{equation}
where $S_{ij} = \braket{\phi^{\mathrm{OTPD}}_i|\phi^{\mathrm{OTPD}}_j}$
are the elements of the overlap matrix and
$b_i = \braket{\Pi|\phi^{\mathrm{OTPD}}_i}$, and
the decomposition error is
\begin{equation}
\Lambda =
\int \Big(\Pi(\vec r) - \sum_i c_i \phi^{\mathrm{OTPD}}(\vec r) \Big)^2 \de^3 \vec r =
\braket{\Pi | \Pi} - \vec b^\dagger \vec S^{-1} \vec b.
\end{equation}
Thus,
to optimize the exponents, we minimize the sum of decomposition errors $\Lambda$
for the molecules considered.

At the first stage, we optimize the exponents shell by shell:
first for $l=0$,
then freeze them and optimize the exponents for $l=1$, etc.
For each angular momentum $l$,
we use a well-tempered Ansatz\cite{Huzinaga1990}
\begin{equation}
    \label{eq:wellT}
    \alpha_i = \beta \alpha_{i-1} \left( 1+ \gamma \left( \frac{i-2}{N+1} \right)^2 \right),\quad i = 2, \ldots, N
\end{equation}
and optimize the parameters $\alpha_0$, $\beta$, and $\gamma$ with the simplex algorithm
as implemented in \texttt{scipy}.\cite{Nelder1965,Wright1996}
The maximum angular momentum $l_\mathrm{max}=4$ and the number of functions $n_l$ for each $l$
are dictated by the composition of the ``reference'' cc-pVTZ-jkfit set (see Table~\ref{tab:basis}).\cite{Weigend2002} With respect to the cc-pVTZ-jkfit set, $\phi^{\mathrm{OTPD}}(\vec r)$ has an additional radial function for each angular momentum, as it improved significantly the final fitting. Addition of further radial functions, or angular momenta did not provide any significant decrease of error (see Table~\ref{tab:basis-errors}),
but introduce more numerical noise and further computational complexity.

\begin{table}[ht]
\caption{\label{tab:basis-errors}{
Absolute (A.E.) and relative decomposition error (R.E.) of the on-top pair density using the optimized basis with $n_l$, $n_l+1$ and $n_l+2$ radial functions. $n_l$ refers to the number of radial functions per angular momentum of the standard cc-pVTZ-jkfit basis.\cite{Weigend2002}
}}
\begin{ruledtabular}
\begin{tabular}{ccccccc}
&\multicolumn{2}{c}{$n_l$}
&\multicolumn{2}{c}{$n_l+1$}
&\multicolumn{2}{c}{$n_l+2$}\\
Molecule     &  A.E. (a.u.) &  R.E. (\%)  & A.E. (a.u.) & R.E. (\%)    & A.E. (a.u.) & R.E. (\%) \\\hline
\ce{H2}      & \num{3.20E-07}         & \num{4.27E-02}       & \num{2.84E-08}        & \num{3.79E-03}        & \num{1.70E-08}        & \num{2.27E-03}     \\
\ce{CH4}     & \num{1.02E-05}         & \num{1.40E-07}       & \num{3.71E-06}        & \num{5.08E-08}        & \num{6.00E-06}        & \num{8.21E-08}     \\
\ce{C2H2}    & \num{1.95E-05}         & \num{1.32E-07}       & \num{4.70E-06}        & \num{3.18E-08}        & \num{1.18E-05}        & \num{8.03E-08}     \\
\ce{C2H4}    & \num{2.72E-05}         & \num{1.85E-07}       & \num{1.17E-05}        & \num{7.94E-08}        & \num{1.22E-05}        & \num{8.32E-08}     \\
\ce{C2H6}    & \num{2.07E-05}         & \num{1.42E-07}       & \num{7.68E-06}        & \num{5.26E-08}        & \num{1.41E-05}        & \num{9.64E-08}     \\
\ce{N2}      & \num{1.35E-04}         & \num{2.06E-07}       & \num{4.54E-05}        & \num{6.95E-08}        & \num{1.35E-05}        & \num{2.06E-08}     \\
\ce{NH3}     & \num{8.91E-05}         & \num{2.79E-07}       & \num{1.45E-05}        & \num{4.56E-08}        & \num{5.44E-06}        & \num{1.70E-08}     \\
\ce{N2H2}    & \num{2.12E-04}         & \num{3.28E-07}       & \num{4.61E-05}        & \num{7.15E-08}        & \num{1.96E-05}        & \num{3.04E-08}     \\
\ce{H2O}     & \num{1.46E-04}         & \num{1.29E-07}       & \num{4.00E-05}        & \num{3.53E-08}        & \num{1.00E-05}        & \num{8.83E-09}     \\
\ce{H2O2}    & \num{3.77E-04}         & \num{1.66E-07}       & \num{1.08E-04}        & \num{4.75E-08}        & \num{3.42E-05}        & \num{1.50E-08}     \\

\end{tabular}
\end{ruledtabular}
\end{table}

At the second stage, the exponents $\{\alpha_\mu\}$ for all the angular momenta
are optimized simultaneously
using the BFGS algorithm
as implemented in \texttt{scipy} (tolerance parameter \num{e-7}).\cite{B1970,F1970,G1970,S1970}
The basis set exponents optimized in the first stage are taken as the initial guess for the BFGS optimization.
The exponents are parameterized as $\alpha_\mu = \exp(p_\mu)$.
The first derivatives of the loss functions~$\Lambda$ with respect to the exponents are computed as follows:
\begin{equation}
\pdd{\Lambda}{\alpha_\mu} =
\vec c^\dagger \left( \pdd{\vec S}{\alpha_\mu}\, \vec c - 2\,\pdd{\vec b}{\alpha_\mu} \right),
\end{equation}
and all the overlap integrals and their derivatives are taken numerically
on the (50, 302) grid for H and (75,302) grid for C, N, O.\cite{Treutler1995} The optimized exponents are available in a separate file, as well as in the Materials Cloud database.
\begin{table}[ht]
\caption{\label{tab:basis}{
Number of radial functions $n_l$ for each angular momentum $l$
in the cc-pVTZ-jkfit and $\phi^{\mathrm{OTPD}}(\vec{r})$ bases.
}}
\begin{ruledtabular}
\begin{tabular}{ccccc}
&\multicolumn{2}{c}{cc-pVTZ-jkfit}&\multicolumn{2}{c}{$\phi^{\mathrm{OTPD}}(\vec{r})$}\\
$l$            & H        & C, N, O  &  H        & C, N, O \\ \hline
0              & 4        & 10       &  5        & 11      \\
1              & 3        & 7        &  4        & 8       \\
2              & 2        & 5        &  3        & 6       \\
3              & 1        & 2        &  2        & 3       \\
4              &          & 1        &           & 2       \\
\end{tabular}
\end{ruledtabular}
\end{table}

\clearpage
\section{Performance of the basis on stretched bonds}

The optimization of the $\phi^{\mathrm{OTPD}}(\vec{r})$ exponents, as detailed in the previous Section, has been performed on representative small molecules at their equilibrium geometry. Nevertheless, the aim of this work is more general and includes molecules with stretched, compressed and even breaking covalent bonds. Therefore, it is necessary, as well, to assess the quality of the decomposition onto $\phi^{\mathrm{OTPD}}(\vec{r})$ for out-of equilibrium configurations.

\begin{figure*}[!htb]
\centering
\includegraphics[width=1.0\textwidth]{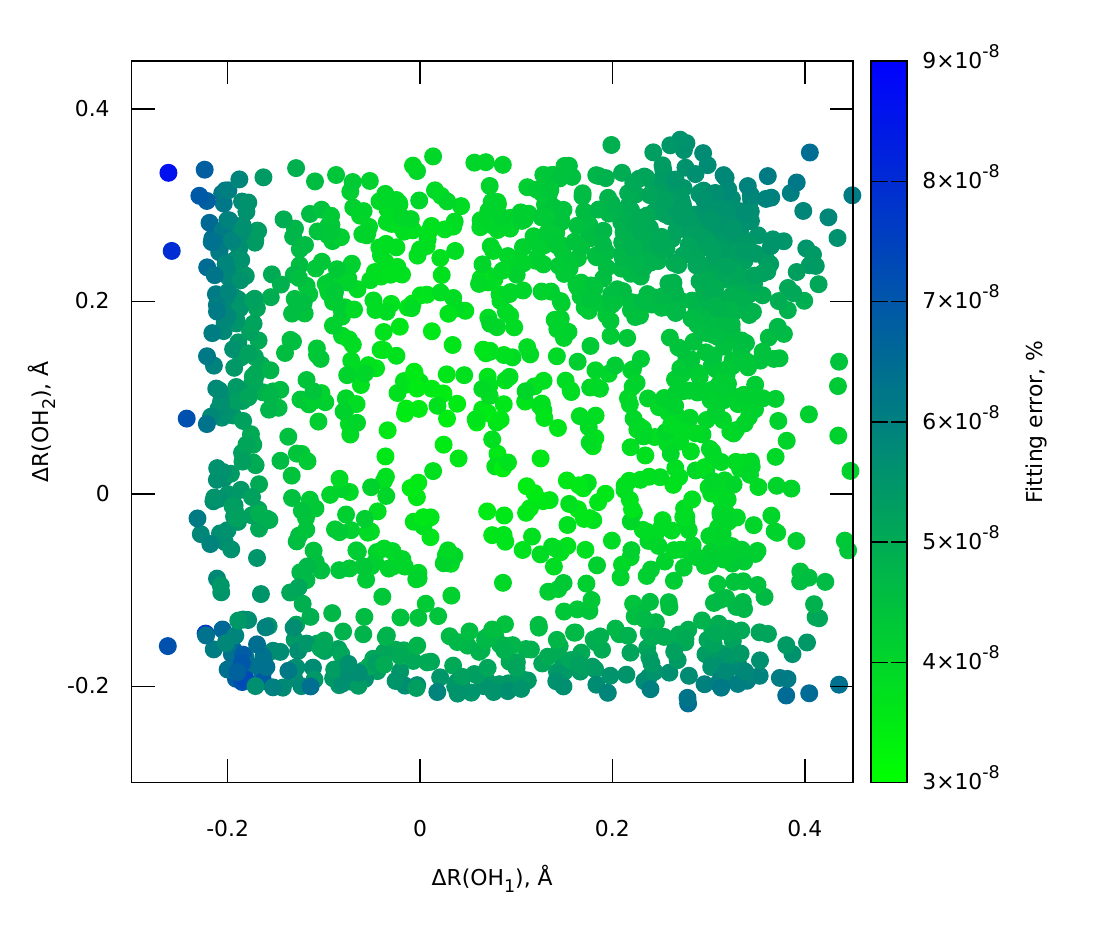}
\caption{
Relative decomposition errors for a set of 1500 water molecules.
The dataset is generated by molecular dynamics simulation ($NVT$~ensemble) at 4000~K
at the GFN2-xTB level. 
The on-top pair density is computed at the HF/cc-pVTZ level.
Bond stretching parameters $\Delta \rm R$(OH) are defined
as difference between the actual bond length and
the bond length for the geometry used for basis set optimization (0.975~\AA).
}
\label{fig:stretched-h2o}
\end{figure*}
\begin{figure*}[!htb]
\centering
\includegraphics[width=0.49\textwidth]{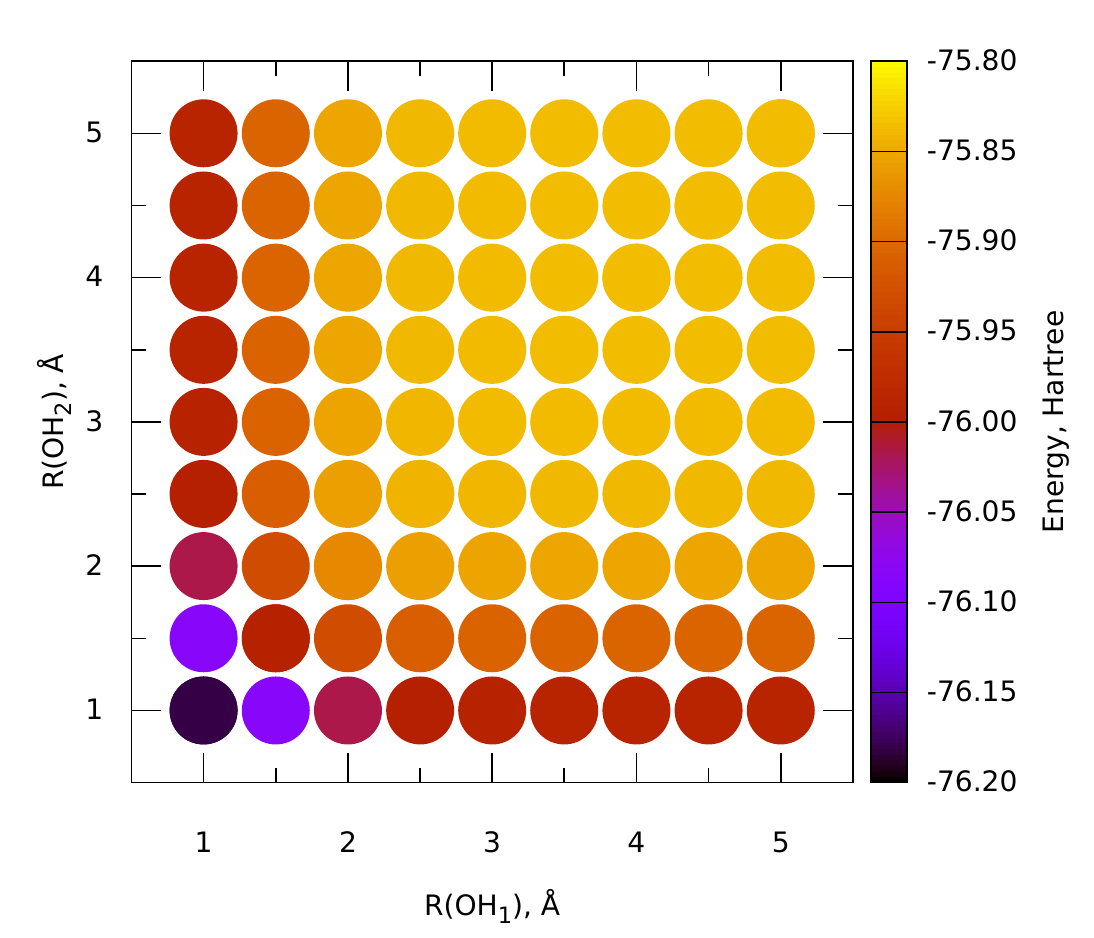}
\includegraphics[width=0.49\textwidth]{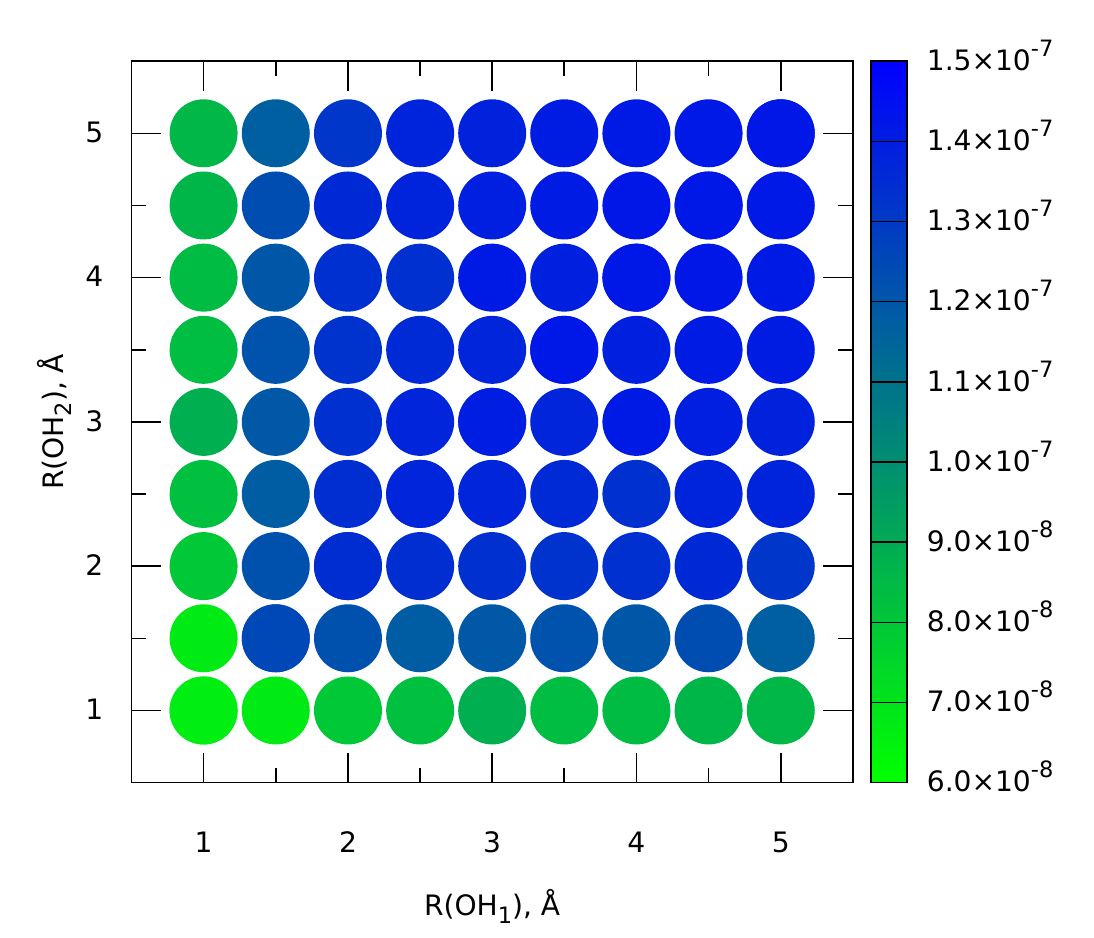}
\caption{\emph{(left)} Potential energy surface of the water molecule at different O--H bond lengths up to 5~\AA. Each row and column represents an asymmetric stretching, while the diagonal of the plot shows the symmetric case. \emph{(right)} Relative decomposition error [\%] of the on-top pair density. Both the energy and $\Pi(\vec{r})$ were computed at CASSCF(8,8)/cc-pVTZ level.}
\label{fig:very-stretched-h2o}
\end{figure*}
\begin{figure*}[!htb]
\centering
\includegraphics[width=0.49\textwidth]{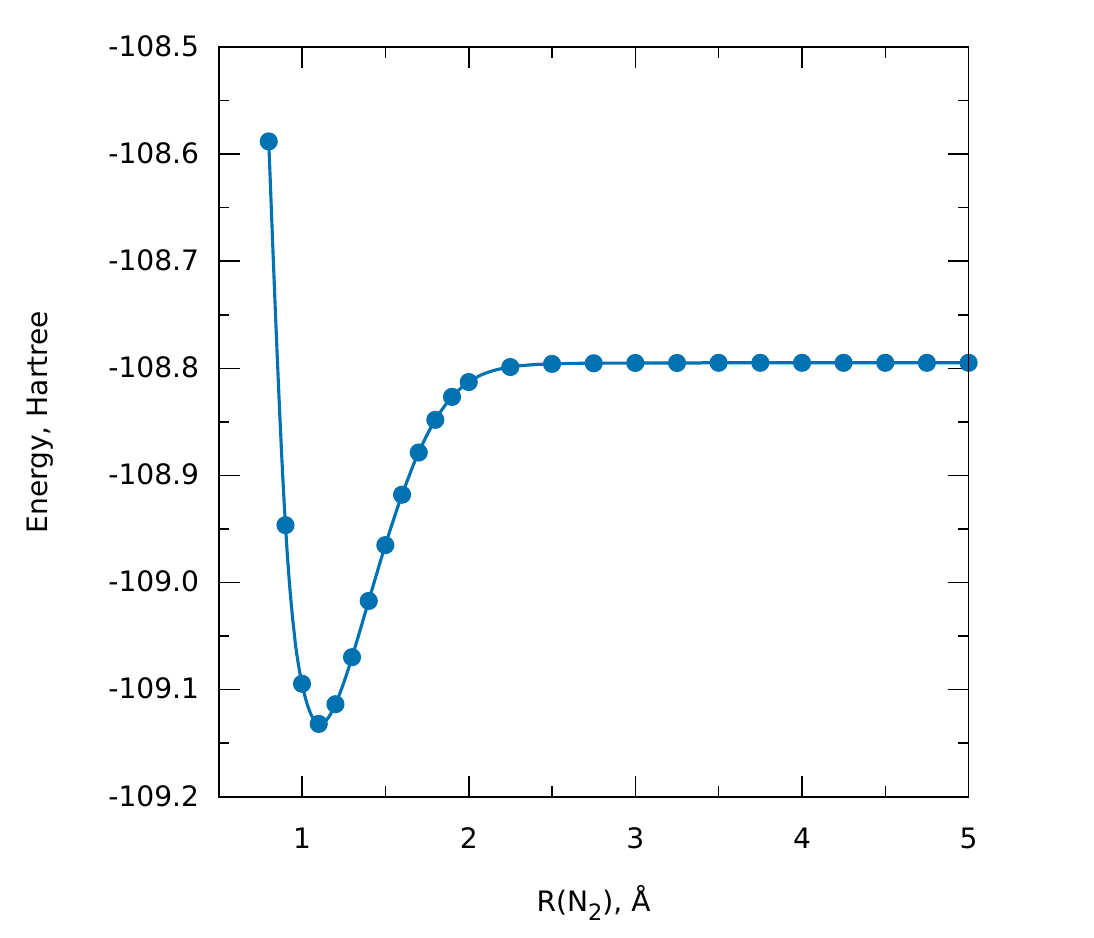}
\includegraphics[width=0.49\textwidth]{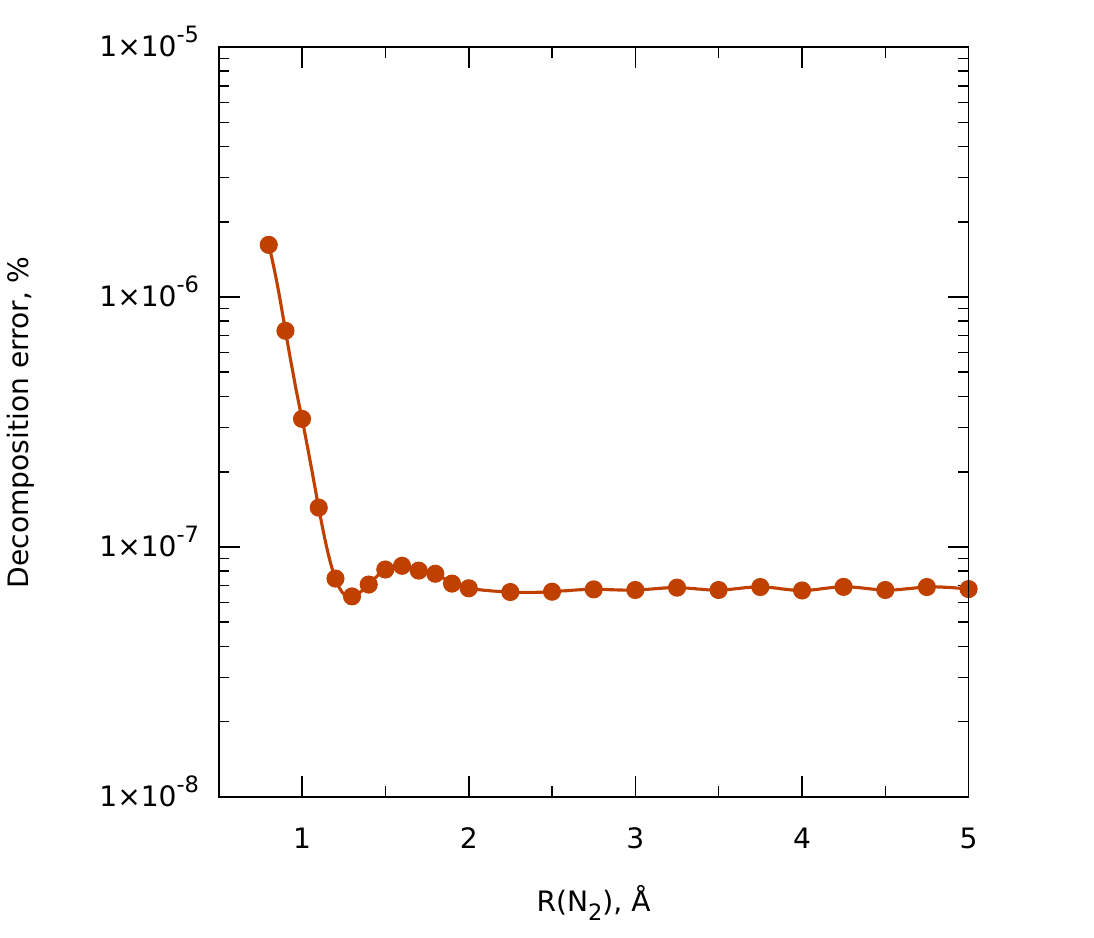}
\caption{
\emph{(left)} Potential energy surface of N2 at increasing N-N bond lengths up to 5~\AA. \emph{(right)} Relative decomposition error [\%] of the on-top pair density. Both the energy and $\Pi(\vec{r})$  were computed at CASSCF(10,8)/cc-pVTZ level.
}
\label{fig:very-stretched-n2}
\end{figure*}

As an illustrative example, Figure~\ref{fig:stretched-h2o} shows the performance of the basis for 1500~different configurations of the water molecule. The configurations were obtained by performing a ground-state molecular dynamics simulation of the water molecule at PBE0/def2-SVP in the $NVT$~ensemble at a temperature of 5000~K.
As shown in the Figure, the error of the decomposition is the lowest for the configurations around equilibrium, as expected from the optimization procedure. Nevertheless, although the fitting error slightly increases in highly stretched configurations, its order of magnitude remains stable and becomes irrelevant when compared to the prediction error (see main text). The same trend is observed on a more systematic scan of the potential energy surface of the water molecule at CASSCF level (symmetric and asymmetric stretching) up to the dissociation limit of the hydrogen atoms (Figure~\ref{fig:very-stretched-h2o}). To complement the results on the water molecule, we demonstrate the robustness of the $\Pi(\vec{r})$ decomposition code for the \ce{N2} molecule at different N--N bond lengths (Figure~\ref{fig:very-stretched-n2}). In this case, the decomposition error becomes nearly constant beyond a distance of 2~\AA, where the on-top pair density of \ce{N2} starts behaving as the one of two independent nitrogen atoms. As in the case of water (see Figure~\ref{fig:stretched-h2o}), the error rises more quickly upon compression of bonds rather than upon their stretching. Nevertheless, even at the most compressed point, the error in the fitting remains orders of magnitudes lower than the prediction error.

\newpage
\section{Decomposition and prediction error with respect to multireference character}

\begin{figure*}[!htb]
\centering
\includegraphics[width=1\textwidth]{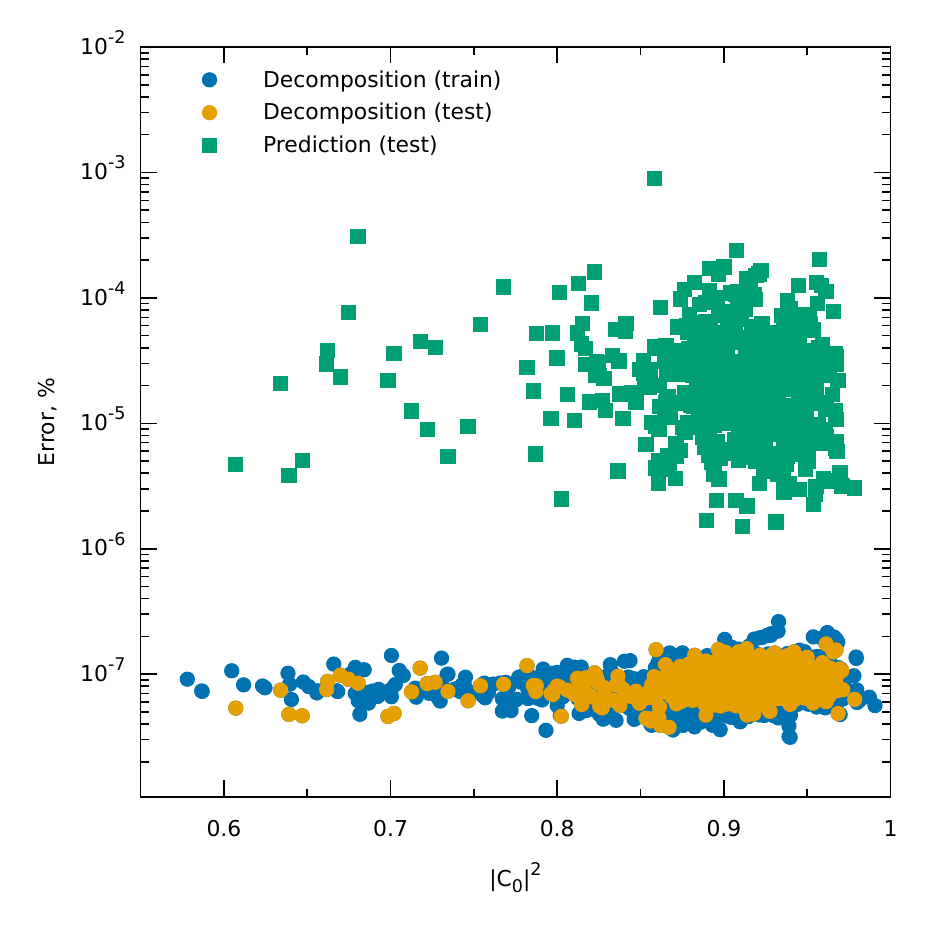}
\caption{
Relative decomposition errors on the training \protect\marksymbol{*}{gnuplot6} and test \protect\marksymbol{*}{gnuplot4} set
and relative prediction errors on the test set \protect\marksymbol{square*}{gnuplot2}
for the on-top pair density
as a function of the weight of the dominant configuration $|C_0|^2$.
}
\label{fig:c0}
\end{figure*}

The model proposed in this work serves as a powerful tool to identify both dynamic and static electron correlation effects in real-space. For this reason, it is crucial to demonstrate that the accuracy of the prediction and basis set decomposition is independent from the multireference character of the molecule under consideration. For this reason, we report in Figure~\ref{fig:c0} a correlation plot between the weight of the dominant electronic configuration ($|C_0|^2$) and the fitting error on both the training and the test set, as well as the prediction error on the test set.

As shown by the blob-like clustering of the errors, the model is not biased by the single- or multireference character of the molecules and no linear correlation between the errors and the molecular $|C_0|^2$ is identifiable.

\newpage

\section{Extrapolation}

\begin{figure*}[!htb]
\centering
\includegraphics[width=0.75\textwidth]{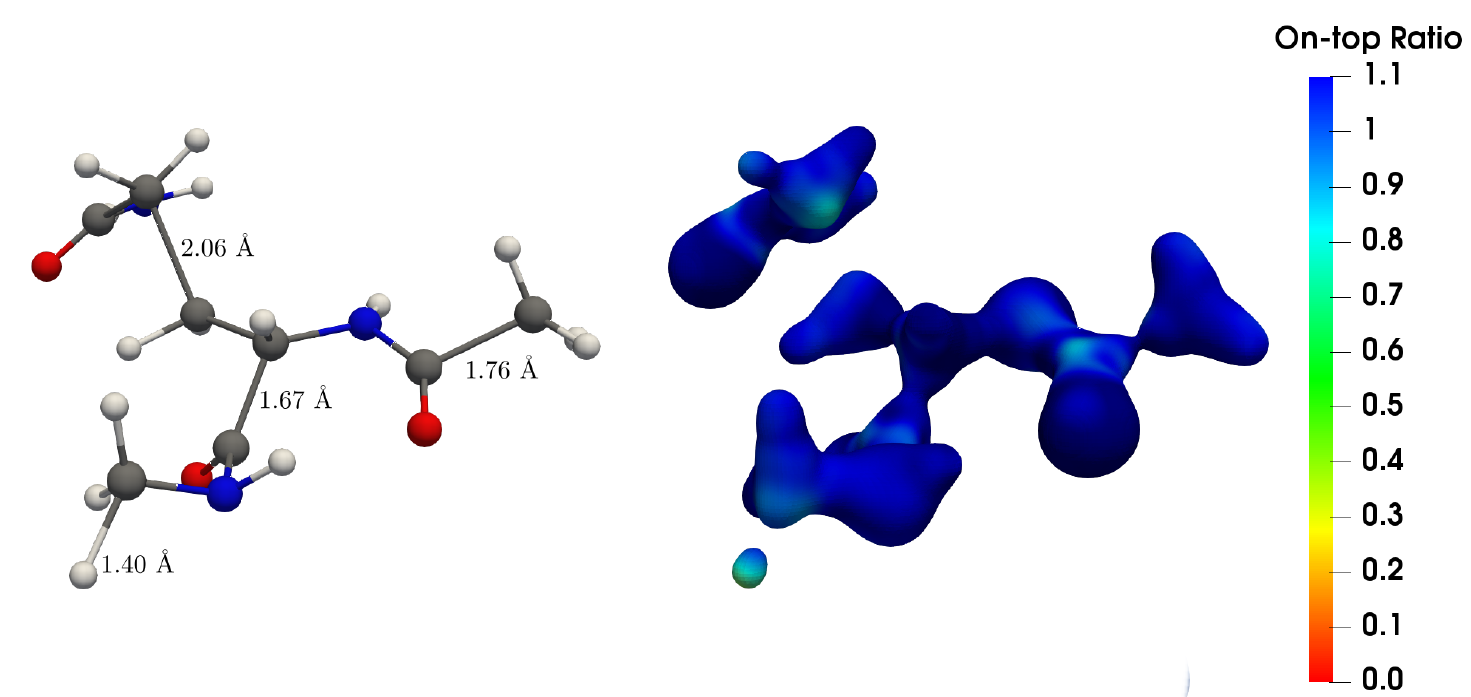}
\caption{Predicted on-top ratio [color-code, $R(\vec{r})$] of a thermally excited dipeptide ($T = 3000$~K) projected on its squared density (isovalue 0.025 $\mathrm{e^2 \cdot Bohr^{-6}}$).}
\label{fig:extrapol}
\end{figure*}

A machine-learning model for chemical applications is transferable when it is able to predict the properties of a larger molecule after training only on its constitutive fragments.\cite{Grisafi2019, Fabrizio-CHEMSCI-2019} The key to transferability is the choice of a local molecular representation (\textit{i.e.} the representation of a molecule in terms of its atom-centred environments, here $\lambda$-SOAP) and the choice of a local target (here $\Pi(\vec{r})$ and $\rho^2(\vec{r})$ decomposed onto an atom-centred basis set).
As a showcase of the transferability of the model proposed in this work, we report in Figure~\ref{fig:extrapol} the on-top ratio of a thermally excited dipeptide, computed from its predicted on-top pair density and density squared. The highly distorted dipeptide geometry was taken from a molecular dynamics simulation ($NVT$~ensemble, $T = 3000$~K).

The Figure shows how the on-top ratio decreases rapidly from the unit in the regions with broken or partially broken bonds. Overall, this result demonstrates the qualitative applicability of the framework on a compound three times larger than the average of the training set.

\newpage
\section{Numerical results}

\begin{table}[h]
\caption{\label{tab:lc}{
Prediction errors on the test set (615 molecules)
for the on-top pair density $\Pi(\vec r)$
and the squared half-density $\frac14 \rho^2(\vec r)$
for different number of molecules in the training set.
}}
\begin{ruledtabular}
\begin{tabular}{ccccc}
&\multicolumn{2}{c}{$\Pi(\vec r)$}&\multicolumn{2}{c}{$\frac14 \rho^2(\vec r)$}\\
Training set size & Absolute error (a.u.) & Relative error (\%) & Absolute error (a.u.) & Relative error (\%) \\ \hline
318  & \num{8.47E-02}   & \num{1.23E-04}        & \num{3.15E-01}        & \num{1.15E-04} \\
637  & \num{3.71E-02}   & \num{5.40E-05}        & \num{1.40E-01}        & \num{5.09E-05} \\
1275 & \num{2.70E-02}   & \num{3.92E-05}        & \num{1.02E-01}        & \num{3.72E-05} \\
1912 & \num{2.48E-02}   & \num{3.60E-05}        & \num{9.38E-02}        & \num{3.41E-05} \\
2550 & \num{2.28E-02}   & \num{3.32E-05}        & \num{8.65E-02}        & \num{3.14E-05} \\
\end{tabular}
\end{ruledtabular}
\end{table}

\bibliography{bibliography}